\documentclass[floats,floatfix,showpacs,amssymb,prd,twocolumn,superscriptaddress,nofootinbib]{revtex4-1}

\usepackage{amssymb,amsmath,verbatim,mathtools,needspace,enumitem,etoolbox,graphicx,physics,microtype,afterpage}

\usepackage[dv	ipsnames, usenames]{xcolor}

\definecolor{linkcolor}{rgb}{0.0,0.3,0.5}
\usepackage[unicode, colorlinks=true, linkcolor=linkcolor, citecolor=linkcolor, filecolor=linkcolor,urlcolor=linkcolor, pdfusetitle]{hyperref}
\usepackage[all]{hypcap}
\usepackage[T1]{fontenc}
\usepackage[utf8]{inputenc}
\usepackage{tabularx}
\allowdisplaybreaks
\newcommand{\ssim}{\mathchar"5218\relax\,}

\def\apj{\rm{ApJ}}
\def\apjl{\rm{ApJ}}
\def\apjs{\rm{ApJS}}
\def\aap{\rm{A\&A}}

\def\mnras{\rm{MNRAS}}
\def\prd{\rm{PRD}}
\def\prl{\rm{PRL}}

\def\lrr{\rm{LRR}}

\newcommand{\jhu}{\affiliation{Department of Physics and Astronomy, Johns Hopkins University, 3400 N. Charles
Street, Baltimore, MD 21218, USA}}
\newcommand{\bham}{\affiliation{School of Physics and Astronomy and Institute for Gravitational Wave Astronomy, University of Birmingham, Birmingham, B15 2TT, UK}}

\begin{document}

\title{Escape speed of stellar clusters from multiple-generation\\black-hole mergers in the upper mass gap}

\author{Davide Gerosa}
\email{d.gerosa@bham.ac.uk}
 \bham
\author{Emanuele Berti}
\email{berti@jhu.edu}
 \jhu


\date{\today}

\begin{abstract}
Pair instabilities in supernovae might prevent the formation of black holes with masses between $\ssim 50 M_\odot$ and $\ssim 130 M_\odot$. Multiple generations of black-hole mergers provide a possible way to populate this ``mass gap'' from below. However this requires an astrophysical environment with a sufficiently large escape speed to retain merger remnants, and prevent them from being ejected by gravitational-wave recoils. We show that, if the mass gap is indeed populated by multiple mergers, the observation of a single black-hole binary component in the mass gap implies that its progenitors grew in an environment with escape speed $v_{\rm esc} \gtrsim 50$ km/s. This is larger than the escape speeds of most globular clusters, requiring denser and heavier environments 
such as nuclear star clusters or disks-assisted migration in galactic nuclei. {A single detection in the upper mass gap would hint at the existence of a much larger population of first-generation events from the same environment, thus providing a tool to disentangle the contribution of different formation channels to the observed merger rate.}
\end{abstract}

\maketitle

\section{Introduction}

The masses of binary black holes (BHs) observed so far by the LIGO and Virgo gravitational-wave (GW) detectors show some evidence for a pileup between $30$ and $45~M_\odot$, as well as a cutoff above $\ssim45~M_\odot$~\cite{ 2018arXiv181112940T}. Sharp cutoff features can be more easily extracted from the data \cite{2017ApJ...851L..25F,2018ApJ...856..173T,2018arXiv180506442W}, thus providing clean opportunities to constrain the underlying astrophysics. Leading explanations for the presence of a mass cutoff at $\ssim45~M_\odot$ are pulsational pair-instability supernovae and pair-instability supernovae in massive stars~\cite{1964ApJS....9..201F,1967PhRvL..18..379B}. These processes are associated with severe mass loss~\cite{2002ApJ...567..532H}, which reduces the masses of BHs formed from stellar collapse. Astrophysical models including pair instabilities predict a sharp drop in the number of detected binary BHs with individual masses $\gtrsim 40$--$50~M_\odot$~\cite{2016A&A...594A..97B,2017MNRAS.470.4739S,2018arXiv181013412M,2019arXiv190402821S}, the precise threshold being quite uncertain (see e.g.~\cite{2017ApJ...836..244W} for a review). 

A few mechanisms have been proposed to fill this ``upper mass gap'' (a ``lower mass gap'' may also exist between the most massive neutron stars and the least massive BHs~\cite{2011ApJ...741..103F}). For instance, older, more massive population III stars might produce more massive remnants, but they are expected to form only a small fraction of all LIGO/Virgo detections~\cite{2014MNRAS.442.2963K,2016MNRAS.460L..74H,2017MNRAS.471.4702B}. Even if stellar collapse limits BHs to masses below $\ssim 45 M_\odot$, sources might appear more massive because of gravitational lensing~\cite{2018PhRvD..98j4029D,2018PhRvD..98j3022C}, but there is no compelling evidence of lensing signatures in the GW signals observed so far~\cite{2019ApJ...874L...2H}.

In this work we will focus on the possibility that the mass gap could be filled by multiple generations of mergers. As BHs merge, they form heavier remnants. If these remnants pair with other BHs and generate GWs, they could potentially provide a detectable population of binaries with one or both component masses in the mass gap. 
Globular clusters and nuclear star clusters have both been invoked to give rise to multiple generations of BH mergers \cite{2009ApJ...692..917M,2017PhRvD..95l4046G,2017ApJ...840L..24F,2016ApJ...824L..12O,2016ApJ...831..187A,2018ApJ...858L...8C,2018ApJ...866...66M,2018MNRAS.481.2168M, 2018PhRvL.120o1101R}. Large gaseous disks surrounding supermassive BHs could also provide favorable environments: stellar-mass BHs embedded in such disks are expected to be affected by migration traps, which provide a natural mechanism to produce multiple mergers~\cite{2017MNRAS.464..946S, 2017ApJ...835..165B,2018MNRAS.474.5672L}. Runaway collisions in clusters have long been considered a leading formation channel for intermediate-mass BHs~\cite{2004ApJ...616..221G,2006ApJ...640..156G,2008ApJ...686..829H,2012MNRAS.425..460M,2016MNRAS.459.3432M,2018PhRvD..97l3003K,2019MNRAS.486.5008A}.

GW signals from multiple-generation BHs might present unique signatures and could potentially be disentangled from the rest of the population \cite{2017PhRvD..95l4046G}. Second-generation BHs are expected to have higher masses -- potentially populating the mass gap -- and a characteristic spin distribution peaked at $\ssim 0.7$ \cite{2008ApJ...684..822B,2017PhRvD..95l4046G,2017ApJ...840L..24F}.  These findings led to the speculation that the heaviest event to date, GW170729, might contain a second-generation BH~\cite{2019arXiv190307813K,2019arXiv190306742C}.

An astrophysical environment can produce multiple-generation GW events only if merger remnants from the previous generations are efficiently retained.  Because of asymmetric GW emission, merger remnants can have substantial recoil speeds (or ``kicks'')~\cite{1961RSPSA.265..109B,1962PhRv..128.2471P,1973ApJ...183..657B,1983MNRAS.203.1049F}. Remnant BHs will be efficiently ejected unless the escape speed of the environment matches typical kicks at merger~\cite{2004ApJ...607L...9M,2007ApJ...667L.133S,2016ApJ...831..187A}. In other words, if multiple-generation BHs populate the mass gap, the escape speed of their birthplace must be sufficiently small. In this paper we quantify this statement, showing that observations of GW events in the mass gap provide a lower limit on the escape speed of their environment.

\section{The model}

Consider a collection of $N$ BHs, which for simplicity we will call a \emph{cluster}. In this toy model, a cluster is simply an  environment with constant escape speed $v_{\rm esc}$: we do not necessarily refer to specific astrophysical settings, such as globular clusters, young star clusters or nuclear star clusters. We randomly choose two BHs from the cluster and we estimate the properties of their merger remnant using fitting formulas to numerical-relativity simulations: we compute the final mass as in Ref.~\cite{2012ApJ...758...63B}, the final spin as in Ref.~\cite{2016ApJ...825L..19H}, and the kick using the expression collected in Ref.~\cite{2016PhRvD..93l4066G} following~\cite{2007ApJ...659L...5C,2007PhRvL..98i1101G,2008PhRvD..77d4028L,2012PhRvD..85h4015L,2013PhRvD..87h4027L,2008PhRvD..77d4028L}. These formulas are evaluated assuming isotropic spin directions.
If the recoil velocity of the remnant BH is greater than $v_{\rm esc}$ the merger product is removed, otherwise it is left in the cluster. We iterate this procedure and randomly extract pairs of BHs until less than two objects remain. %

First-generation BHs are  injected into the cluster with  masses distributed according to $p(m)\propto m^{\gamma}$, and dimensionless spin magnitudes distributed  uniformly in $[0,\chi_{\rm max}]$. Crucially, we assume the presence of a mass gap and restrict the initial BH masses to $m\in[5,50]M_\odot$. We choose $\gamma=-2.3$ as expected from the Kroupa initial mass function \cite{2001MNRAS.322..231K} and used to model GW event rates \cite{2018arXiv181112907T}.

The pairing probability (i.e., how BHs in the cluster choose their partners) is a key ingredient of the model. We explore two possibilities:
\begin{itemize}
\item \emph{Random pairing}. We randomly pick two BHs from the cluster with uniform probability, i.e. $p_{\rm pair}(m_1,m_2)=$ const.
\item \emph{Selective pairing}. We assume a pairing probability {$p(m_1)\propto m_1^\alpha$, $p(m_2|m_1)\propto m_2^\beta$ with $m_1>m_2$, $\alpha=-1.6$, and $\beta=6.7$}
as measured by Ref.~\cite{2018arXiv181112940T} using current GW data (cf. their model B).
\end{itemize}
Selective pairing favors mergers with $m_1\simeq m_2$, as predicted for mass-segregated clusters and currently supported by the data \cite{2019arXiv190512669F}. {We opt for a pairing prescription motivated by current observations, but alternative expressions have been derived theoretically (see e.g.~\cite{2016ApJ...824L..12O}).}

Every time two BHs merge, their GW signal could potentially be detected. We pair each merger with a redshift value $z$ extracted uniformly in comoving volume and source-frame time, i.e. $p(z)\propto (dV_c/dz)/ (1+z)$.
This is a simple prescription to average over a large number of clusters at different redshifts.
 We then estimate the probability of detection $p_{\rm det}(m_1,m_2,z)$ as in Refs.~\cite{1993PhRvD..47.2198F,1996PhRvD..53.2878F,2010ApJ...716..615O,2015ApJ...806..263D,2018PhRvD..98h3017T,2016ApJ...819..108B,2017arXiv170908079C,gwdet}. We set a signal-to-noise-ratio threshold of $\rho_{\rm thr}=8$~\cite{2016ApJ...833L...1A}, computed assuming a single LIGO instrument at design sensitivity~\cite{2018LRR....21....3A} and the waveform model of Ref.~\cite{2016PhRvD..93d4007K} for nonspinning
BHs (Refs.~\cite{2015ApJ...806..263D,2018PhRvD..98h4036G,2018PhRvD..98h3007N} showed that spins have a marginal effect on $p_{\rm det}$ unless strong alignment is present, which is not our case).
The probability $p_{\rm det}$ corresponds to the (unnormalized) detection rate, allowing us to estimate detector selection effects on the GW events resulting from our model.

For each cluster, we define $p_{\rm gap}$ to be the fraction of component masses in the mass gap weighted by the 
detection rate $p_{\rm det}$.

\section{Results}
\label{results}

To summarize, our toy model has four main parameters: 
\begin{itemize}
\item the escape speed of the cluster $v_{\rm esc}$,
\item the largest injected spin $\chi_{\rm max}$,
\item a pairing prescription, \emph{random} or \emph{selective},
\item the initial number of objects $N$. 
\end{itemize} 
For each choice of these parameters, we simulate several clusters to decrease counting errors.

\begin{figure}
\includegraphics[width=\columnwidth]{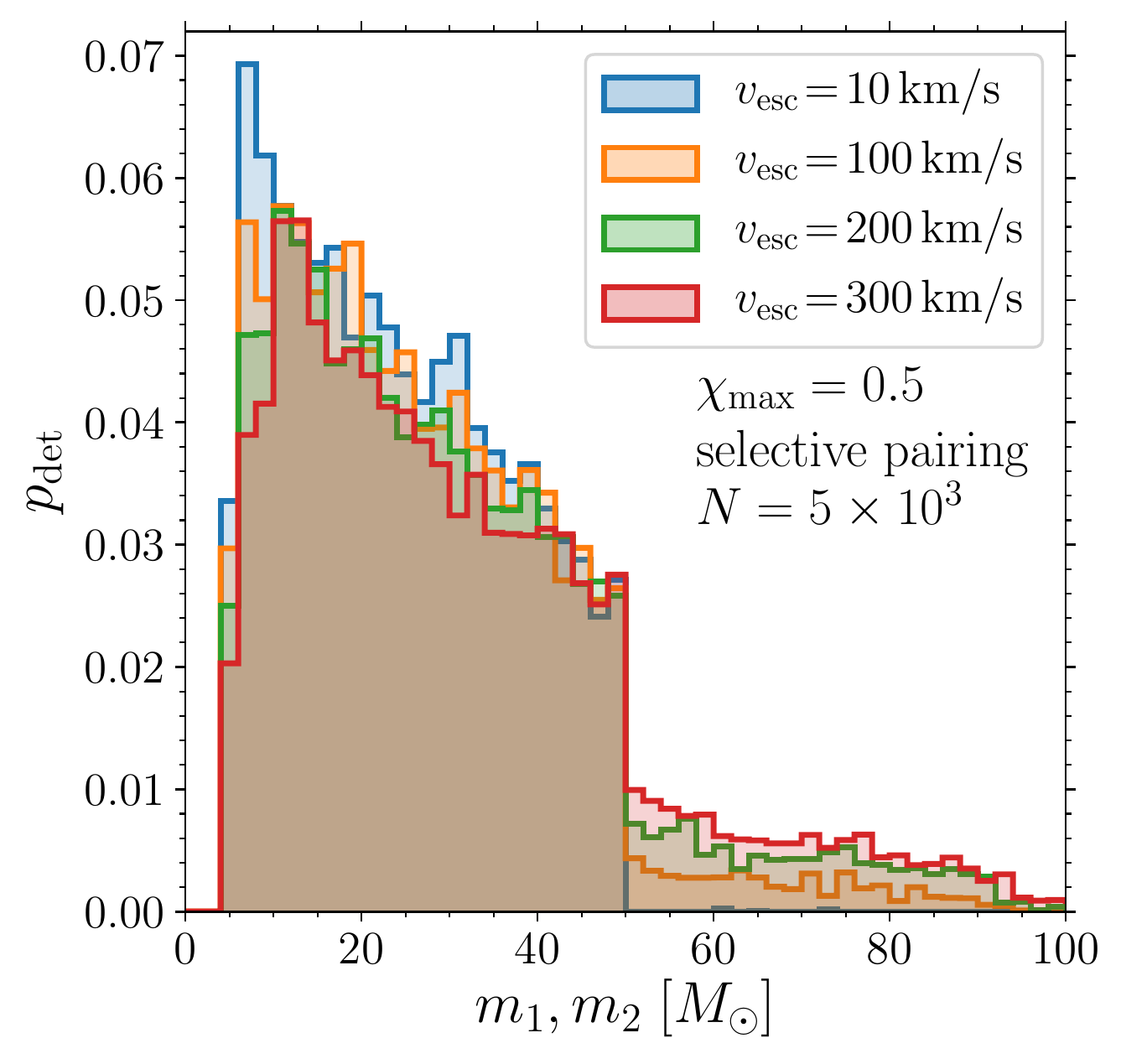}
\caption{Detectable distribution of component masses for clusters with different escape speeds. If the escape speed is large enough, second-generation mergers populate the mass gap. We assume $\chi_{\rm max}=0.5$, $N=10^3$, and selective pairing.}
\label{Mvesc}
\end{figure}

\begin{figure*}
\centering
\includegraphics[height=0.98\columnwidth]{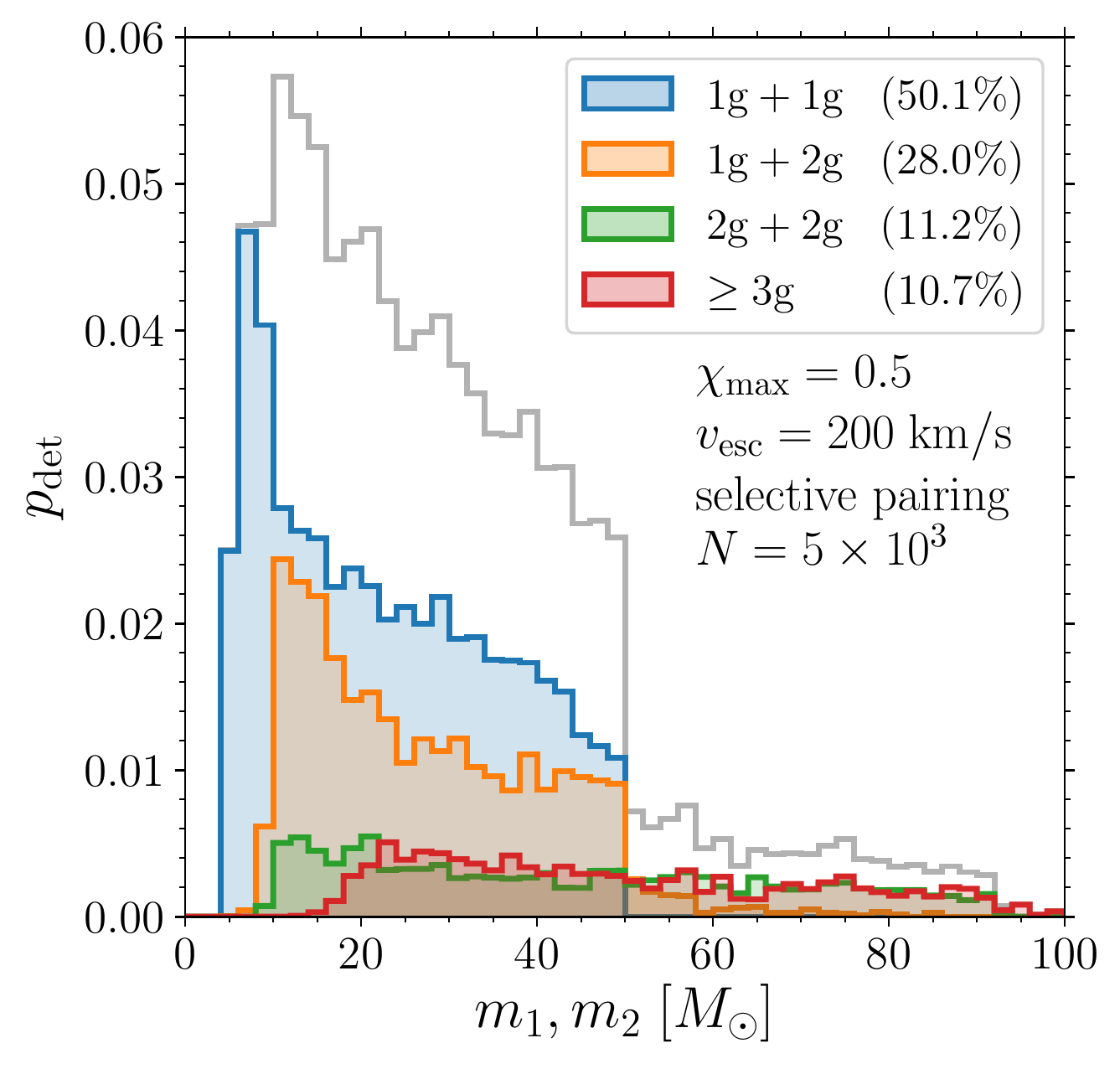}
\includegraphics[height=0.98\columnwidth]{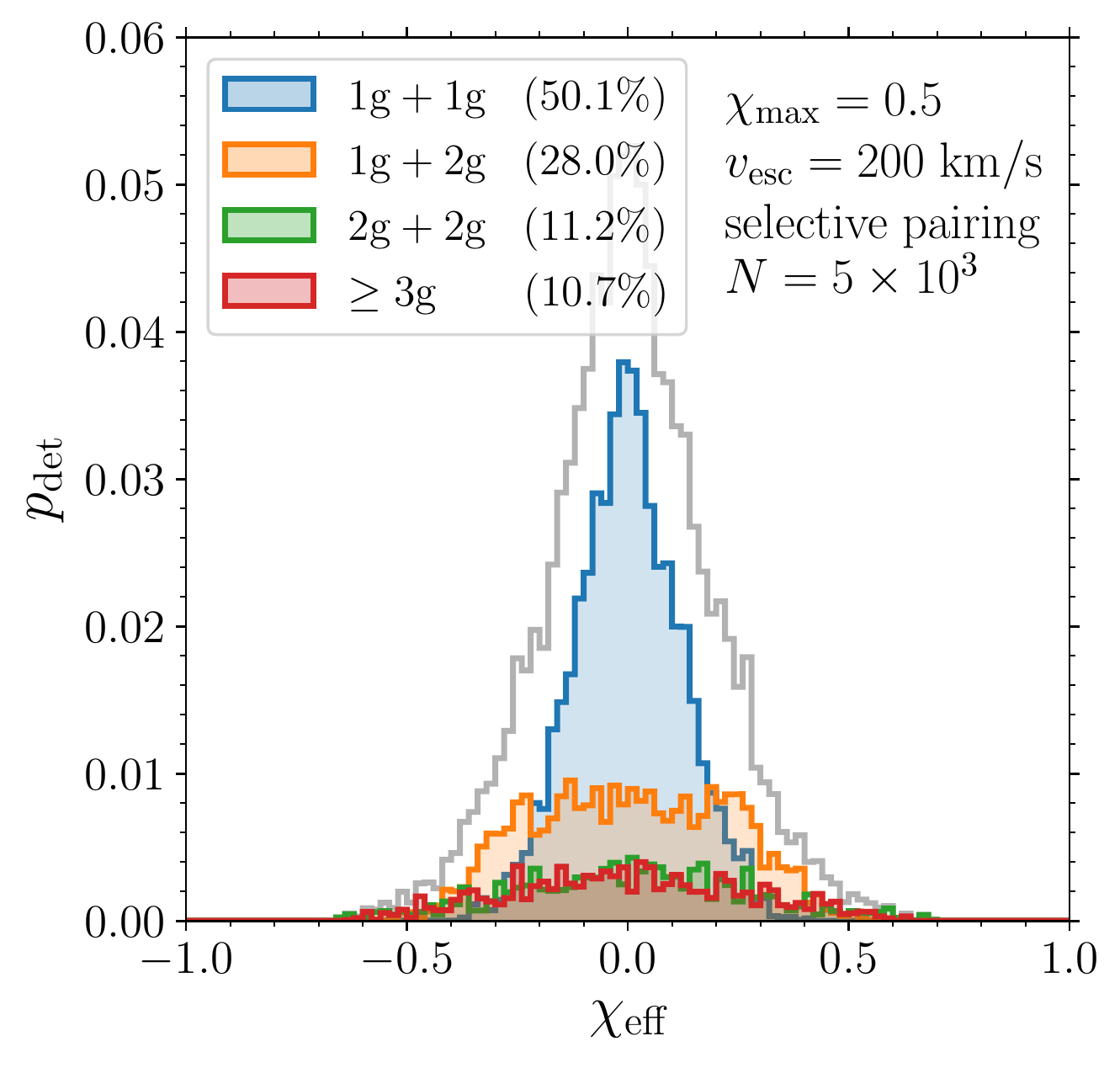}
\caption{Component masses (left)  and effective spins (right) of detectable GW events in each BH merger generation for a representative model with $v_{\rm esc}=200$ km/s, $\chi_{\rm max}=0.5$, $N=10^3$, and selective pairing. Colors refer to merger between first-generation BHs (1g), second-generation BHs (2g), or any higher generation ($\geq$ 3g). The contribution of  each class to the detection rate is specified in the legend. The empty gray histogram (sum of the filled colored ones) shows the entire distribution.}
\label{gencontent}
\end{figure*}

Figure~\ref{Mvesc} shows the detectable distribution of component masses for some of our models. Multiple generations of mergers allow BHs to leak into the mass gap. As expected, clusters with higher escape speeds retain more BHs and populate the mass gap more efficiently. Besides the number of systems in the gap, the escape speed affects also the slope of the mass spectrum in that region: larger (smaller) values of $v_{\rm esc}$ yield steeper (shallower) spectra. In principle, a very large number of observations could allow us  to measure the slope, hence $v_{\rm esc}$.
On the other hand, the shape of the mass spectrum below the mass gap ($5M_\odot<m_1,m_2< 50M_\odot$ ) is only mildly dependent on the escape velocity $v_{\rm esc}$.

\begin{figure}[h!]
\includegraphics[width=0.99\columnwidth]{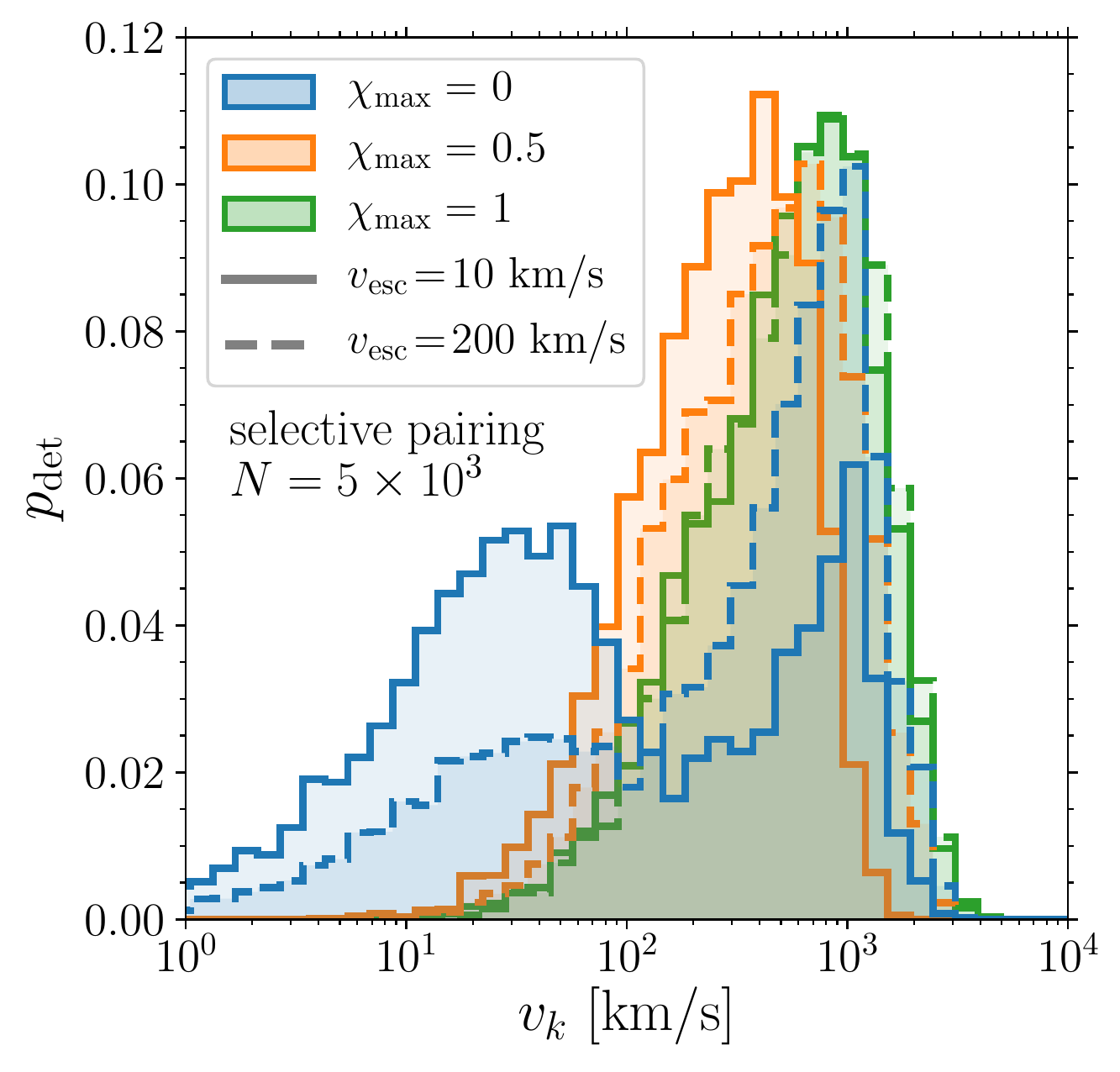}
\caption{Distribution of recoil velocities for detectable binaries in clusters with different escape speeds (solid/dashed linestyles) and maximum injected spins (colors). We fix $N=10^3$ and consider selective pairing. Distributions generically peak at $v_k\ssim500$ km/s, unless spins are very low: in this case the distribution is bimodal, with a second peak at $v_k\ssim 30$~km/s.}
\label{kickchimax}
\end{figure}

In Fig.~\ref{gencontent} we analyze the contribution of different generations of merging BHs to the mass and spin distributions in a representative case ($v_{\rm esc}=200$~km/s, $\chi_{\rm max}=0.5$, and selective pairing). We find that $\ssim 50\%$ of the BHs belong the first generation of mergers (1g+1g). 
The fraction of BH binaries where only one of the two components had a previous merger (1g+2g) is $\ssim 30 \%$ of the detectable population, but these BHs do not appreciably fill the mass gap.
On the contrary, roughly half of the 2g+2g population is found in the mass gap.
This a direct consequence of selective pairing: large BHs from subsequent generations are more likely to pair with other heavy merger remnants. About $10\%$ of the component masses results from mergers of generation higher than two ($\geq$ 3g). 
Component masses $>100 M_{\odot}$ (i.e. twice the assumed cutoff) contribute to only $\lesssim 0.5\%$ of the detection rate.

Spin measurements are a distinguishing feature of multiple-generation mergers, because merger remnants are, on average, rapidly rotating.  As shown in Fig.~\ref{gencontent}, binaries with at least one merger remnant (1g+2g, 2g+2g, $\geq$ 3g) do indeed have a broader distribution for the so-called effective spin $\chi_{\rm eff}$
~\cite{2001PhRvD..64l4013D,2008PhRvD..78d4021R,2015PhRvD..92f4016G}. The symmetry of the effective-spin distributions about $\chi_{\rm eff}=0$ reflects our assumption that spin directions are isotropic before merger, as expected in dense environments, where GW sources form dynamically.

\begin{figure*}
\includegraphics[page=1,width=\columnwidth]{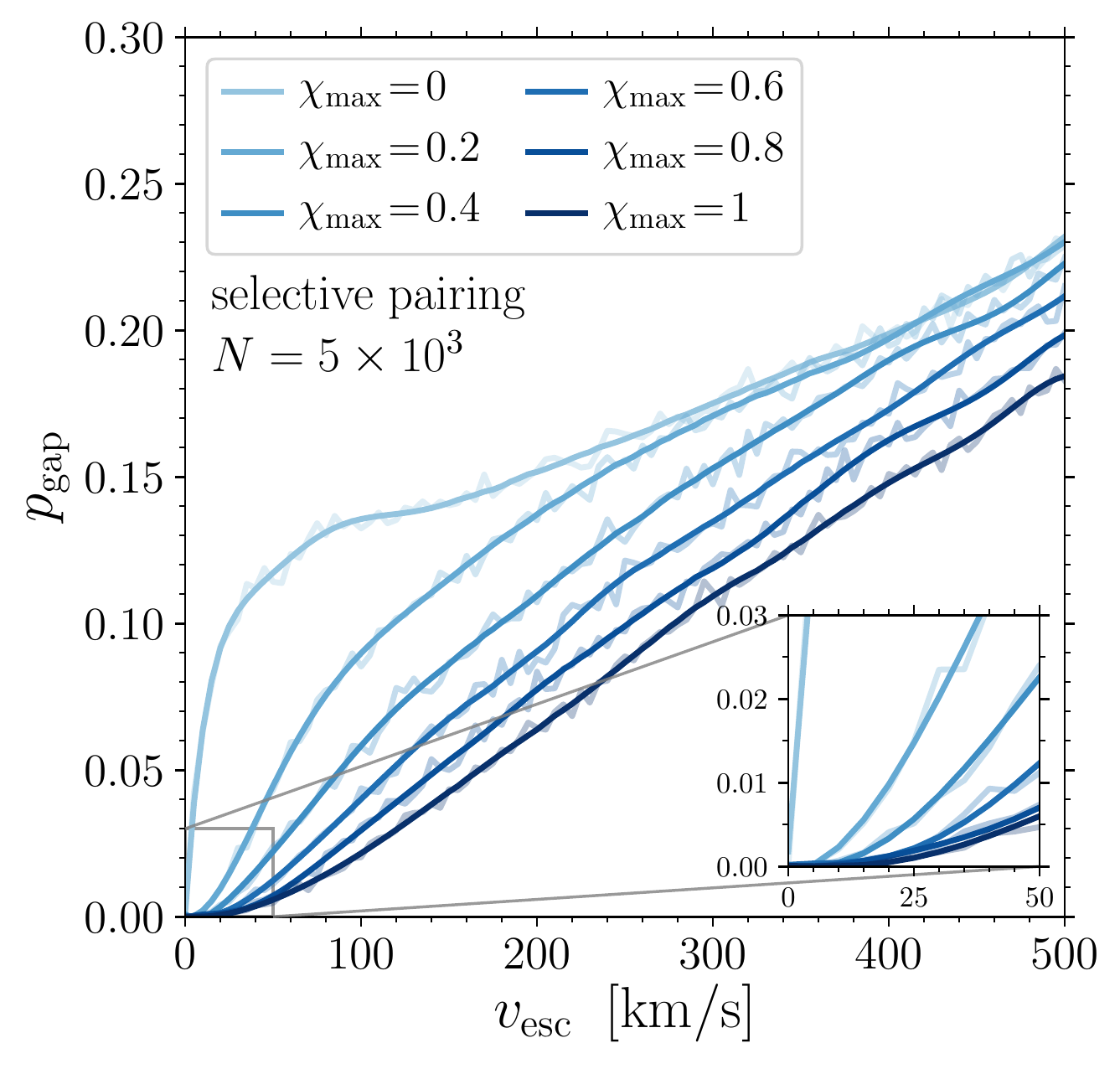}
\includegraphics[page=2,width=\columnwidth]{gapesc}
\caption{Probability $p_{\rm gap}$ of detecting a BH in the mass gap in clusters with different escape speed $v_{\rm esc}$. Left (right) panel corresponds to the case of selective (random) paring. Colors indicate different injected spins, from $\chi_{\rm max}=0$ (light) to $\chi_{\rm max}=1$ (dark). The size of the cluster is set to $N=10^3$. Light curves show  results of our runs; thick curves show a polynomial fit. }
\label{gapesc}
\end{figure*}

The probability of retaining BHs inside clusters, hence filling the mass gap, depends on the magnitude of BH kicks. Figure~\ref{kickchimax} shows recoil distributions weighted by the detection rate. For generic values of the BH spins, the kick distribution peaks at $\ssim 500$~km/s, and second-generation BHs are easily ejected from clusters with low escape speed. %
If instead the spins of first-generation BHs are fine-tuned to be very small ($\chi_{\rm max}\simeq 0$), the distribution of $v_k$ is bimodal. The remnants of first-generation mergers receive kicks of $\ssim 30$~km/s and are easily retained.
Second-generation BHs have larger spins, and receive kicks of %
 $\ssim 10^3$~km/s. This is the typical value expected for merger remnants: for $q=1$ and $\chi_1=\chi_2=0.7$, the fitting formula of Refs.~\cite{2016PhRvD..93l4066G,2007ApJ...659L...5C,2007PhRvL..98i1101G,2008PhRvD..77d4028L,2012PhRvD..85h4015L,2013PhRvD..87h4027L,2008PhRvD..77d4028L} returns $120\,{\rm km/s} \lesssim v_k\lesssim 2190 \,{\rm km/s}$ in 90\% of the cases, with a median of $\ssim 765$ km/s. The relative contribution of the low- and high-kick modes of the distribution depends on the escape speed:
if $v_{\rm esc}$ is small (large), low (high) kicks are more likely.

Figure~\ref{gapesc} shows our main result: a single BH binary merger observation with a component BH in the mass gap places a lower limit on the escape speed of its environment.

The probability of detecting BHs in the mass gap is strongly correlated with $v_{\rm esc}$.  For random pairing, we find that $p_{\rm gap}\lesssim 2\%$ for $v_{\rm esc} \lesssim 100$ km/s. 
The constraints are less stringent for the (presumably more realistic) case of selective pairing: we find $p_{\rm gap}\lesssim 2\%$ for $v_{\rm esc}\lesssim 50$ km/s, unless spins are fine-tuned to be very small. This is because selective pairing implies that $m_1\simeq m_2$, so the remnants of first-generation mergers receive kicks $\mathcal{O}(10)$~km/s. For $\chi_{\rm max}\simeq 0$ and selective pairing, most second-generation BHs are retained in the cluster and populate the mass gap, even if $v_{\rm esc}$ is very low. However, this is a rather fine-tuned scenario: already with spin magnitudes uniformly distributed between $0$ and $\chi_{\rm max}=0.2$ we find $p_{\rm gap}\lesssim 2\%$ for $v_{\rm k}\lesssim 30$ km/s. Note that all of our assumptions on the spin magnitude are compatible with current observations, even for large values of $\chi_{\rm max}$: the distributions are always strongly peaked at $\chi_{\rm eff}=0$ and symmetric about $\chi_{\rm eff}=0$ (cf. Fig.~\ref{gencontent}).

\begin{figure}
\includegraphics[width=\columnwidth]{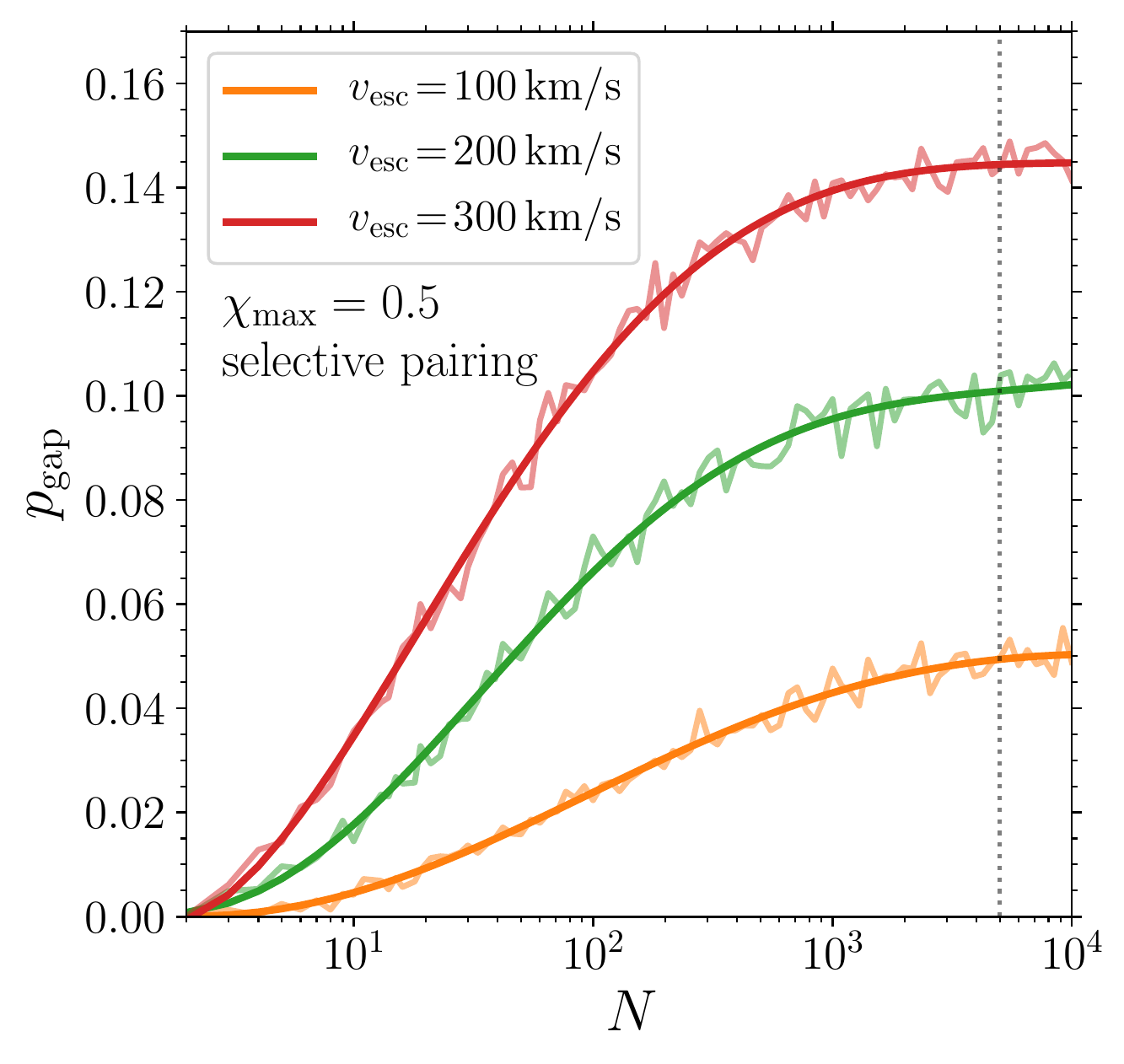}
\caption{Probability $p_{\rm gap}$ of detecting a BH in the mass gap as a function of the number $N$ of BHs in the cluster. The probability $p_{\rm gap}$  increases roughly monotonically with $N$ and tends to a constant for $N\gtrsim 10^3$. %
Orange, green, and red curves are computed assuming $v_{\rm esc}=100$, $200$, and $300$~km/s, respectively. We set $\chi_{\rm max}=0.5$ and assume selective pairing. Light curves show the results of our runs; thick curves show a polynomial fit. The vertical dotted line marks the value of $N$ used in all other figures of this paper.}
\label{gapN}
\end{figure}

Finally, Fig.~\ref{gapN} shows that the probability $p_{\rm gap}$ of detecting a BH in the mass gap converges very rapidly with the number of BHs in the cluster.   As expected, $p_{\rm gap}$ is small for very small values of $N$ ($p_{\rm gap}=0$ in the limit $N=2$), but it is roughly constant for $N\gtrsim 10^3$. All of our results assume $N=5\times 10^3$, and in this sense they can be regarded as conservative.

\section{Discussion}

We use a simple model to relate the fraction of GW events with component BH masses in the mass gap to the escape speed of their astrophysical environment. Assuming that stellar collapse does indeed produce a BH mass spectrum with an upper cutoff,  we find that only environments with low escape speeds allow for multiple mergers to populate the mass gap. In particular, a single GW observation with $M\gtrsim50 M_\odot$ would point towards an astrophysical environment with $v_{\rm esc}\gtrsim 50$~km/s. 
This lower limit exceeds the escape speed of most globular clusters~\cite{2004ApJ...607L...9M,2016ApJ...831..187A},
requiring environments with larger escape speeds (such as galactic nuclei).

The mass gap can be more efficiently populated if first-generation BHs have small or zero spin. This is consistent with, e.g., the model of Ref.~\cite{2018PhRvL.120o1101R}, where all BHs formed from stellar collapse are assumed to be nonspinning, and found to be retained in some globulars. However, this is a rather fine-tuned scenario: we find that even a small fraction of spins with magnitude $\chi\ssim 0.2$ is enough to efficiently deplete clusters. 

{Although our simple model cannot predict absolute merger rates, we can predict what fraction of the total number of detections is in the mass gap (Fig.~\ref{gapesc}). A single detection of a BH with $M\gtrsim50 M_\odot$ would indicate that a much larger population of objects born in a very dense environment should be present in the data. This might allow us to disentangle the contributions of different formation channels to the observed merger rate.}

The simplicity of our model requires care in interpreting our results. First, we assumed environments with constant escape speed, while in reality the ejection probability depends on the merger location inside the cluster. However BHs are (on average) heavier than other stellar populations, and by mass segregation they should preferentially populate the cluster's central region, where a constant escape speed is a reasonable approximation. 
Second, we only considered ejections due to BH recoils. Dynamical interactions and natal kicks may be more efficient than gravitational recoils at expelling BHs from clusters. In this sense, our lower limit on $v_{\rm esc}$ is conservative.
Third, we neglect time delays between the various merger generations and the redshift evolution of the cluster. Even if merger remnants are retained, clusters may not have time to assemble multiple generations of mergers. This is also a conservative assumption, which lowers $p_{\rm gap}$. Time delays might help us to distinguish populations of mergers in the observed catalog \cite{2017PhRvD..95l4046G}, as higher-generation BHs will be found at lower redshifts.

The pairing probability is perhaps the most crucial ingredient of our model. We implemented a prescription for selective pairing, where BHs are more likely to form binaries with companions of similar masses. This scenario is well motivated by mass segregation in cluster dynamics, and tentatively favored by GW data~\cite{2019arXiv190512669F,2018arXiv181112940T}. Selective pairing turns out to be a conservative assumption when estimating the escape speed, lowering the limit on $v_{\rm esc}$ by about a factor of 2 compared to the more simplistic assumption of random pairing. %

{More work is needed to generalize our simple model to realistic astrophysical settings. Requiring that merging BHs form in environments with $v_{\rm esc}\gtrsim 50$~km/s might have important consequences on other astrophysical properties, including the stellar density, the rate of dynamical interactions, the  response to orbital perturbations~\cite{2018MNRAS.474.5672L}, and the fraction of eccentric mergers~\cite{2019ApJ...871..178G}.}

In conclusion, detections of BH mergers with component masses $\gtrsim 50 M_\odot$ challenge our current understanding of stellar collapse in massive stars. If explained by the assembly of two or more smaller BHs, such GW events would provide a conservative lower limit on the escape speed of the merger's astrophysical environment.

\acknowledgments
We thank A.~Vecchio, A.~Klein, C.~Moore, S.~Toonen, T.~Woods, M.~Volonteri, B.~Kocsis, and the anonymous referee for comments and discussions.  E.B. is supported by NSF Grant No. PHY-1841464, NSF Grant No. AST-1841358, NSF-XSEDE Grant No. PHY-090003, and NASA ATP Grant No. 17-ATP17-0225. Computational work was performed on the University of Birmingham's BlueBEAR cluster and at the Maryland Advanced Research Computing Center (MARCC).

\bibliography{escape}

\end{document}